%% file: Bossini_etal.tex
\documentclass[fleqn,usenatbib]{mnras}

\usepackage{mathptmx}

\usepackage[T1]{fontenc}
\usepackage{ae,aecompl}

\usepackage{graphicx}	
\usepackage{amsmath}	
\usepackage{amssymb}	
\usepackage[usenames,dvipsnames]{color} 

\usepackage{mathrsfs}
\usepackage{verbatim}
\usepackage[normalem]{ulem}

\input{journalDefs}

\newcommand{\Mmc}{\mbox{$M_\mathrm{mc}$}}	 
\newcommand{\Mcc}{\mbox{$M_\mathrm{cc}$}}	 
\newcommand{\Teff}{\mbox{$T_\mathrm{eff}$}}	
\newcommand{\gradT}{\mbox{$\nabla_T$}}	
\newcommand{\gradr}{\mbox{$\nabla_{\mathrm{r}}$}}
\newcommand{\grada}{\mbox{$\nabla_{\mathrm{ad}}$}}
\newcommand{\gradmu}{\mbox{$\nabla_\mu$}}
\newcommand{\logL}{\mbox{$\log L/\mathrm{L}_{\odot}$}}	
	
\newcommand{\Msun}{\mbox{M$_{\odot}$}}
\newcommand{\dnu}{\mbox{$\Delta\nu$}}		
\newcommand{\numax}{\mbox{$\nu_\mathrm{max}$}}   
\newcommand{\dpg}{\mbox{$\Delta\Pi_\mathrm{1}$}}
\newcommand{\co}{\mbox{C$/$O}}			
\newcommand{\amlt}{\mbox{$\alpha_{\mbox{\tiny\rm mlt}}$}}		
\newcommand{\aovh}{\mbox{$\alpha_{\mbox{\tiny\rm ovH}}$}}		
\newcommand{\aovhe}{\mbox{$\alpha_{\mbox{\tiny\rm ovHe}}$}}
\newcommand{\trialfa}{\mbox{triple-$\alpha$}}
\newcommand{\coreaction}{\mbox{$^\mathrm{12}$C($\alpha$,$\gamma$)$^\mathrm{16}$O}}

\newcommand{\schw}{Schwarzschild}	
\newcommand{\kepler}{{\it Kepler}}

\newcommand{\beq}{\begin{equation}}		
\newcommand{\eeq}{\end{equation}}

\title[Uncertainties on near-core mixing in RC stars] {Uncertainties on near-core mixing in red-clump stars: effects on the period spacing and on the luminosity of the AGB bump.}

\author[Bossini et al.]{
Diego Bossini$^{1,2}$\thanks{E-mail: dbossini@bison.ph.bham.ac.uk},
Andrea Miglio$^{1,2}$, 
Maurizio Salaris$^{3}$, 
Adriano Pietrinferni$^{4}$, \newauthor                        
Josefina Montalb\'an$^{5}$, 
Alessandro Bressan$^{6}$, 
Arlette Noels$^{7}$, 
Santi Cassisi$^{4}$, \newauthor
L\'eo Girardi$^{8}$, 
and Paola Marigo$^{5}$
\\
  $^{1}$ School of Physics and Astronomy, University of Birmingham, Edgbaston, Birmingham B15 2TT, United Kingdom \\
  $^{2}$ Stellar Astrophysics Centre, Department of Physics and Astronomy, Aarhus University, Aarhus, DK\\
  $^{3}$ Astrophysics Research Institute, Liverpool John Moores University, IC2, Liverpool Science Park, 146 Brownlow Hill,\\
  Liverpool L3 5RF, United Kingdom\\
  $^{4}$ Osservatorio Astronomico di Collurania -- INAF, via M. Maggini, 64100 Teramo, Italy\\
  $^{5}$ Dipartimento di Fisica e Astronomia, Universit\`a di Padova, Vicolo dell'Osservatorio 3, I-35122 Padova, Italy \\
  $^{6}$ SISSA, via Bonomea 265, I-34136 Trieste, Italy\\ 
  $^{7}$ Institut d'Astrophysique et G\'eophysique de l'Universit\'e de Li\`ege, All\'ee du six Ao\^ut, 17 B-4000 Liege, Belgium\\
  $^{8}$ Osservatorio Astronomico di Padova -- INAF, Vicolo dell'Osservatorio 5, I-35122 Padova, Italy \\
 }

\date{Accepted 2015 July 27.  Received 2015 July 27; in original form 2015 May 1}

\pubyear{2015}

\begin{document}
\label{firstpage}

\pagerange{\pageref{firstpage}--\pageref{lastpage}}
\maketitle

\begin{abstract}
Low-mass stars in the He-core-burning phase (HeCB) play a major role in stellar, galactic, and extragalactic 
astrophysics. The ability to predict accurately the properties of these stars, however, depends on our understanding of convection, 
which remains one of the key open questions in stellar modelling. 
We argue that the combination of the luminosity of the AGB bump (AGBb) and the period spacing of gravity modes (\dpg) during the HeCB 
phase, provides us with a decisive test to discriminate between competing models of these stars. We use the MESA, BaSTI, and PARSEC 
stellar evolution codes to model a typical giant star observed by \kepler. We explore how various near-core-mixing scenarios 
affect the predictions of the above-mentioned constraints, and we find that \dpg\ depends strongly on the prescription adopted. Moreover 
we show that the detailed behaviour of \dpg\ shows the signature of sharp variations in the Brunt-V\"ais\"al\"a frequency, which could 
potentially give additional information about near-core features. 
We find evidence for the AGBb among \kepler\ targets, and a first comparison with observations shows that, even if standard models are 
able to reproduce the luminosity distribution, no standard model can account for satisfactorily the period spacing of HeCB stars. 
Our analysis allows us to outline a candidate model to describe simultaneously the two observed distributions: a model with a moderate 
overshooting region characterized by an adiabatic thermal stratification. This prescription will be tested in the future on 
cluster stars, to limit possible observational biases. 
\end{abstract}




\section{Introduction}
\label{sec:intro}
The helium-core-burning (HeCB) phase of low-mass stars affects many aspects of their subsequent evolution, and has therefore been the subject of numerous studies in the literature.
In particular, the red clump (RC) is a well-known feature in the Hertzsprung-Russell diagram -- as well as the color-magnitude diagram --  of simple and composite stellar populations, and is associated to the low-mass, metal-rich stars in the He-core-burning phase \citep{Cannon_70}. 
The RC plays a key role in many fields of astrophysics: its luminosity, for instance, can be used as a distance and age indicator of  clusters and nearby galaxies, while the observed chemical composition of its members is useful to investigate the chemical evolution of galaxies \citep[e.g., see][]{Girardi_Salaris01,Catelan09,Nidever_etal14}.

Current models of the internal structure and evolution of such stars, however, still suffer from systematic uncertainties which are due in most cases to our limited understanding of the physical processes in stellar conditions (energy transport and nuclear processes). In particular, predictions of stellar lifetimes in the HeCB phase are strongly dependent on the (poorly constrained) amount of mixing applied beyond boundaries of convective regions and on the adopted definition (and correct implementation) of such boundaries (e.g. see \citealt{Chiosi07, Castellani_etal71a, Gabriel_etal14, Bressan_etal15}).  Crucially, our ability to test models has been limited so far by the lack of observational constraints which are specific to the internal structure of evolved stars. 

Asteroseismology of thousands of red giants observed by CoRoT \citep{Baglin_etal06} and \kepler\  \citep{Borucki_etal10} has changed the situation. We can now use the pulsation frequencies to place tight constraints not only on the fundamental stellar properties, but also to probe their internal structure (see e.g. \citealt{ChaplinMiglio_13} and references therein). In particular, as presented in \citet{Montalban_etal13}, the frequencies of oscillation modes detected in HeCB stars are sensitive diagnostics of the chemical and thermal stratification of the energy-generating core, providing us with a novel and independent constraint, which is specific to the core structure of HeCB stars.  
In this paper we argue that seismic constraints, when used in conjunction with well-known classical indicators (in particular the luminosity of the AGB bump, see below), provide us with a decisive test to discriminate competing models of different near-core mixing schemes adopted during the He-burning phase.
We also show  how current standard models (see Sec. \ref{sec:models}) from different stellar codes do not match these constraints simultaneously.

The paper is organised as follows: 
we start by reviewing currently available indicators of core mixing in He-core-burning stars (Sect. \ref{sec:review}). We describe in Sec. \ref{sec:models} the set of stellar models we use, while in Sec. \ref{sec:impact} we explore the impact of several commonly adopted prescriptions for near-core mixing on the observables, with emphasis on asteroseismic constraints (Sec. \ref{sec:sismo}). We then compare in Sec. \ref{sec:data} our predictions with asteroseismic constraints as published in recent papers  \citep{Mosser_etal14, APOKASC}. Finally, we present a summary and future prospects in Sec. \ref{sec:conclusion}.

\section{Observational constraints on HE-core-burning models}
\label{sec:review}
One of the  main observables used to constrain the mass of the fully mixed core during the HeCB phase is the $R_{\rm 2}$ {ratio} \citep{Buonanno_etal85}. $R_{\rm 2}$ is defined as the ratio between the number of early Asymptotic Giant Branch (eAGB) Horizontal Branch (HB) stars in simple stellar populations (chemically homogeneous and coeval stars) and is directly connected to the lifetime of the two phases ($R_2\sim\tau_\mathrm{AGB}/\tau_\mathrm{HB}$).
The value of $R_{\rm 2}$ is affected by the core mixing during the HeCB phase, as discussed in, e.g., \citet{Bressan_etal86}, \citet{Caputo_etal89}.
The use of $R_{\rm 2}$ has hitherto been limited to stars in clusters, due to the small number of field stars with accurate 
distance available, and due to the inherent complication of dealing with an ensemble of stars with a spread in age and chemical 
composition (which also hampers robust inferences on their evolutionary state).

Another important observable related to the HeCB and AGB evolution of  low- and intermediate-mass stars is the luminosity of the AGB bump (AGBb, see e.g. the review by \citealt{Catelan07}).  
Similarly to the bump in RGB, the AGBb is a feature of the eAGB that consists in three passages of the evolutionary track in a small interval of luminosity and causes a local increment of stars in the luminosity distribution of a stellar population.
The occurrence of the AGB bump is connected to the formation of the helium-burning shell (see e.g. \citealt{CS13}, paragraph 5.2). 
The luminosity at which it occurs depends on the location in mass of the He shell at its ignition, hence it is determined by the maximum extension of the mixed core during the HeCB phase.
While the AGB bump had been highlighted in stellar evolutionary tracks a few decades ago \citep[see, for instance][]{Caputo_etal78}, its first identification as a distinct observational feature in galaxies was reported by \citet{Gallart98}. 
The AGBb is nowadays observed in a large number clusters and nearby galaxies \citep[e.g., see][]{Alcock_etal97, Ferraro_etal99, Beccari_etal06, Dalcanton_etal12}. 
A further important property of the AGBb is the weak dependence of the luminosity ratio between the RC and the AGBb on the metallicity and initial helium abundance (\citealt{Castellani_etal91}; \citealt{Bono_etal95}).


Asteroseismology of white dwarfs may also provide us with observational constraints to test models in the HeCB and AGB phase.
The C/O  profile of a white dwarf at the beginning of the cooling sequence corresponds to that of the stellar core at the end of the AGB, hence it is largely determined by the combined action of chemical mixing and nuclear burning during the HeCB phase.
Using pulsation modes detected in WDs, \citet{Metcalfe_etal02} found a discrepancy in the central oxygen-to-carbon ratio between stellar models and the value inferred from seismic data. They ascribed this difference to an underestimation of \coreaction\ cross section, however, as pointed out by \citet{Straniero_etal03}, the final \co\ ratio in models also depends on the amount of mixing applied in the convective regions beyond the \schw\ border, and on the adopted definition of such boundaries. 

While providing valuable information, currently available tests cannot be used to discriminate competing models. In this paper we propose a way forward, which is based on the combination of said constraints and the more direct diagnostics of the conditions in the core 
provided by non-radial modes observed in HeCB stars.

\section{Stellar models}
\label{sec:models}

In our exploratory analysis we consider models with \mbox{$M=1.5\,\Msun$} and solar chemical composition. This set of parameters are both typical of the sample of RC giants observed by \kepler\, \citep[see e.g.][]{APOKASC, Mosser_etal14}, and similar to those of giants in the cluster NGC6819 \citep[see][]{Basu_etal11, Miglio_etal12, Sandquist_etal13}.  
Models in HeCB phase are computed using three different stellar structure and evolution codes (MESA, BaSTI, and PARSEC) and several assumptions about near-core mixing. 

A first set of  models presented in this study is computed using the {\it Modules for Experiments in Stellar Astrophysics} code \citep[MESA,][]{Paxton_etal11}.
The choice of parameters and of the relevant physics is the following: 
\begin{itemize}
\item the initial mass is $1.5\,\Msun$, and no mass loss is considered during the RGB;
\item the mass fraction of heavy elements  ($Z=0.01756$), the initial helium abundance ($Y=0.26556$), and the  mixing-length parameter ($\amlt=1.69278$) are those resulting from a solar-calibrated model without diffusion;  
\item the partition of heavy elements is taken from \citet{gn93};
\item the nuclear reaction rates are taken from \citet{NACRE};
\item during the He-burning stage one has to properly account for the C and O abundance changes in the stellar matter and their effect on radiative opacities, hence we use opacity tables accounting for various levels of C and O enhancements (Type 2 opacity tables, \citealt{opal}); 
\item the evolution includes the pre-main sequence phase and is computed until the first thermal pulse along the AGB (TP-AGB);
\item in models computed with overshooting, the overshooting scheme adopted is a step function in which the parameter \aovhe\ indicates the extension of the mixed region from the \schw\ border in units of the local pressure scale-height. The overshooting region is instantaneously mixed \citep{Maeder75}; 
\item we consider two scenarios for the temperature gradient (\gradT) in the overshooting region, according to the definition given in \citet{Zahn91}: \gradT=\gradr\ ({\it overshooting}) or \gradT=\grada\ ({\it penetrative convection}), where \gradr\, and \grada\, follow the usual notation for the radiative and adiabatic temperature gradient;
\item during the main-sequence phase we adopt an overshooting parameter \aovh=0.2 (Nevertheless, the choice of \aovh\ does not impact on the structure of a low-mass HeCB star).
\end{itemize}

\begin{figure}
\centering
\resizebox{\hsize}{!}{\includegraphics{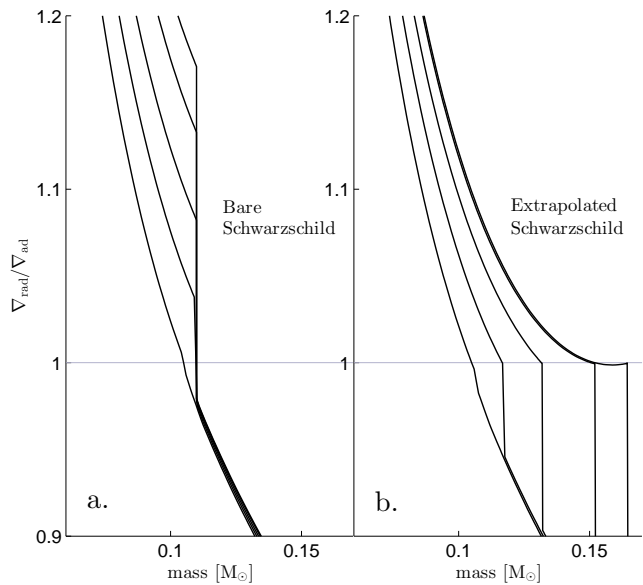}}
\caption{Ratio between the radiative and adiabatic temperature gradients as a function of mass in  MESA HeCB models computed using the BS ({\it panel a}) and ES ({\it panel b}) scheme. The latter leads to a self-consistent definition of the convective boundary (following the \schw\ criterion), as the adiabatic and radiative gradients are equal at the boundary of the core. From bottom to top the lines are for a sequence of models at the stages of $Y_c=0.93$, $0.87$, $0.81$, $0.74$, and $0.70$.}
\label{fig:Schw_crit}
\end{figure}

We compute the following MESA models with different schemes for convective mixing:
\begin{enumerate}
\item[\bf BS] The ``bare \schw'' model (BS, to follow the notation in \citealt{Straniero_etal03}).
In the BS model, the \schw\ criterion is applied on the radiative side of the convective boundary which implies that the convective core is not allowed to grow, leading  to a discontinuity in the chemical composition (hence in the radiative gradient) at the  border of the convective core (Figure \ref{fig:Schw_crit} panel a).  
\end{enumerate}

A recent paper by \citet{Gabriel_etal14} showed how this implementation of the \schw\ criterion leads to an inconsistent location of the convective border, as the convective luminosity is non-zero there ($\gradr > \grada$ at the inner side of the convective boundary). 
The book of \citet{S58} and the analysis presented in \citet{Castellani_etal71a} also led to similar conclusions. The latter consider the BS convective border as in an unstable equilibrium, in the sense that by extending outwards the convective core by an arbitrarily small quantity,  \gradr\ at the new border may be larger than \grada, hence, the region is convectively unstable according to the \schw\ criterion. To find a stable border, the convective core must be extended until radiative and adiabatic gradient become equal ({\sl induced overshooting}). 
However, as the HeCB proceeds (${Y_{\rm c}} \lesssim 0.69$) this scheme generates a local minimum in \gradr\ within the mixed core. 
The addition of more radiative layers surrounding the increasingly  
larger convective core will decrease \gradr\ to the value of \grada\ at the location of the minimum. This creates a separate convective region in layers beyond the location of 
this minimum, where \gradr\ is still larger than \grada. The treatment of this external convective region is problematic. Full mixing between layers inside the minimum 
of \gradr\ and the external convective shell cannot happen, because otherwise the minimum of \gradr\ would decrease below the local value of \grada\, with the contradiction 
of a fully mixed region where \gradr\ is however locally lower than \grada. A solution to this problem is the formation of a partially mixed  --{\sl semiconvective}-- 
region between the minimum of \gradr\ and the outer radiative zone \citep{Castellani_etal71b}. This is  
usually treated with dedicated algorithms that allow for partial chemical mixing to satisfy $\gradr=\grada$ in this region,  
with the consequence of creating a smooth gradient of chemical composition before the sharp discontinuity due to the HeCB.
Although we discourage to adopt the bare \schw\ model, we use this mixing scheme in order to compare our models with those described in the 
literature \citep[e.g.][]{Straniero_etal03, KWW}. 
\begin{enumerate}
\item[\bf HOV]  High overshooting model with $\aovhe=1$ (HOV). 

\item[\bf HPC]  Penetrative Convection model with a high overshooting parameter $\aovhe=1$ (HPC).
\end{enumerate}

For low and moderate values of \aovhe, overshooting models (as well as penetrative convection) might develop a semiconvective region similar to the case presented by \citet{Castellani_etal71b}. 
For high values of the overshooting parameter (e.g HOV and HPC) the extra mixed region becomes large enough to prevent the formation of a semiconvective region \citep[see e.g. ][]{Bressan_etal86, Straniero_etal03}. Moreover, in these cases the determination of the fully mixed region requires the application of the \schw\, criterion in layers where there is no chemical composition gradient/discontinuity, which greatly simplifies the numerical implementation of the convective-instability criterion. 

In order to extend our study of the mixing-schemes commonly adopted, we also consider stellar models from other evolution codes (BaSTI and PARSEC).
To compute those models we used, where possible, the same physical inputs adopted in MESA. 
\begin{enumerate}
\item[\bf  BaSTI-SC]  
We have also computed  models with the BaSTI code \citep[A Bag of Stellar Tracks and Isochrones,][]{Pietrinferni_etal04,Pietrinferni_etal06,Pietrinferni_etal13}. 
The BaSTI model has  $M=1.5\rm M_\odot$, computed for $Z=0.0176$, $Y=0.266$, and $\amlt=1.69$ with  \citet{gn93} heavy element partition.
The input physics relevant to this analysis is the same as in the MESA calculations but for the ${\rm ^{12}C(\alpha,\gamma)^{16}O}$ reaction rate that comes from \citet{Kunz_etal02}. 
Core mixing during the HeCB stage, induced overshooting and semiconvection have been taken into account by adopting the numerical scheme firstly introduced by \citet{Castellani_etal85} and previously described. 
Indeed, a semiconvective region starts to appear when the central abundance by mass of He is $\sim0.7$. 
The occurrence of breathing pulses - which appear when the central abundance of He drops below of $0.12$ - is inhibited by imposing that the abundance of Helium in the core is not allowed to increase at each time step. 
The evolution starts from the pre-main sequence to the RGB-tip, and is resumed at the start of the HeCB 
after the core electron degeneracy has been removed (the helium-flash evolution is 
not computed), and 3\% of carbon has been produced during the He-flash.
No core convective overshoot during the central H-burning stage and no mass loss during the RGB phase are taken in account. 

\item[\bf PARSEC-LOV] 
We also computed models using PARSEC \citep[PAdova \& TRieste Stellar Evolution Code,][]{Bressan_etal12, Bressan_etal13} for $M = 1.5M_\odot$, $Z=0.017$, and $Y=0.279$. 
We briefly summarize below the main input physics adopted for the current models.
The nuclear reaction rates and corresponding $Q$-values are the recommended values in the JINA reaclib database \citep{JINA2}. 
The high-temperature opacities, $4.2 \le \log(T/{\rm K}) \le 8.7$, are provided by the Opacity Project At Livermore (OPAL) team \citep[][and references therein]{opal} and the low-temperature opacities, $3.2 \le \log(T/{\rm K}) \le 4.1$, are from \AE SOPUS\footnote{http://stev.oapd.inaf.it/aesopus} tool \citep{MarigoAringer_09}. 
The equation of state is computed with the FreeEOS code (A.W.~Irwin\footnote{http://freeeos.sourceforge.net/}). 
The heavy element partition is from \citet{Caffau_etal11}.
The MLT parameter, $\amlt=1.74$, is calibrated on the solar model accounting for element diffusion.
The standard mixing scheme in PARSEC considers an overshooting parameter of $\Lambda_c=0.5$ across the formal Schwarzschild border, which means about $0.25H_p$ above it, and a radiative thermal stratification of the extra-mixed region. 
Similar to BaSTI, the evolution starts from the pre-main sequence, stops at the helium flash and restarts after the core electron degeneracy has been removed after the flash. 
The amount of carbon consumed to remove the electron degeneracy is computed from the variation of the gravitational binding energy of the core during the flash.
During the HeCB phase, besides accounting for core overshooting, the code may deal with residual semi-convective instabilities using the \schw\ criterion, and  suppresses possible breathing pulses of convection.

\item[\bf PARSEC-MPC] 
We use PARSEC to compute a modified track in which we extend the mixed He core by penetrative convection (adiabatic stratification) with a moderate value of overshooting ($\Lambda_c=1.0$, i.e. $\aovhe\sim0.5$).

\item[\bf  ES] Finally, we include in the analysis a model obtained with MESA in which we include an external routine to implement a similar prescription as in BaSTI-SC. This routine consists in the following steps:

\begin{enumerate}
\item[1.] at the beginning of each timestep we set the position of the convective-core boundary and fully mix the convective region according to the \schw\ criterion as implemented in MESA.
\item[2.] we let the code calculate the burning during the timestep. 
\item[3.] at the end of the timestep, we check whether the radiative gradient (with the new composition determined by the burning) 
at the convective border is higher than the adiabatic gradient. 
\item[4a.] if this is the case, we restart the timestep from point 1, but extending the core boundary by one mesh.
\item[4b.] if not, the equality of the gradients has been achieved and the code can continue to the next timestep (point 1).
\end{enumerate}
As a result of this algorithm the convective core extends naturally (Fig. \ref{fig:Schw_crit} panel b). We call this model Extrapolated \schw\ (ES). The internal structure and seismic predictions of ES (see Sec. \ref{sec:DPg} and Figures \ref{fig:DP-hec} and \ref{fig:astero}) are remarkably similar to the BaSTI-SC. 
Differently from BaSTI and PARSEC, however, a routine to treat semiconvection is still to be developed in MESA, hence we stop the evolution of the ES model at $Y_{\rm c} \sim 0.69$.
\end{enumerate}

\section{Predicted stellar properties of models in the HeCB and AGB phase}
\label{sec:impact}
We now compare the properties of the series of models presented in Section \ref{sec:models}, with particular emphasis on those which can be tested via a direct comparisons to observations. We start by comparing predictions of non-seismic observables (see Sec. \ref{sec:review}), while in Sec. \ref{sec:sismo} we focus on seismic diagnostics.
   
\begin{figure}
\centering
\resizebox{\hsize}{!}{\includegraphics{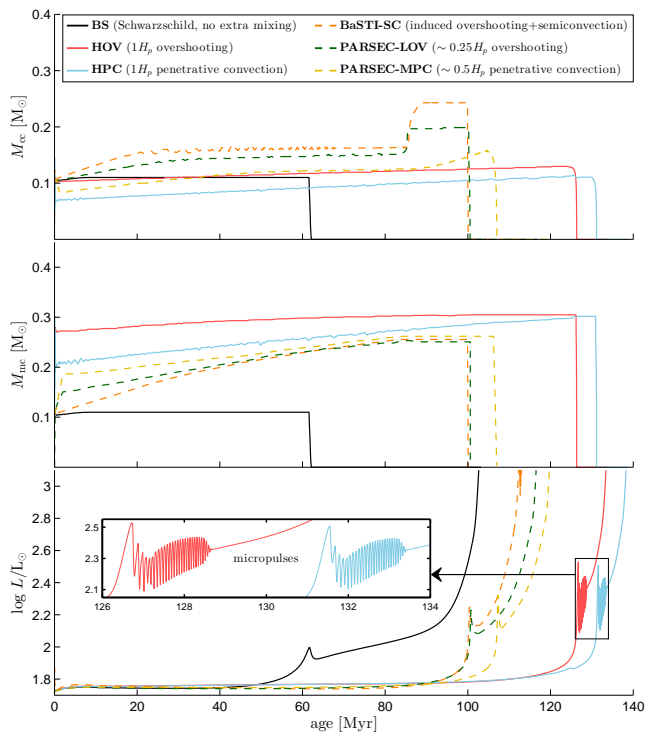}}
\caption{Convective-core (upper panel), mixed-core mass (middle panel) and total luminosity (lower panel) as a function of time for the BS, HOV, HPC, BaSTI-SC, PARSEC-LOV, and PARSEC-MPC models from the start of the HeCB phase up to the first AGB-TP. Both HPC and HOV models show micropulses.}
\label{fig:M-L_age}
\end{figure}

\label{sec:mass}
The main consequence of applying different mixing schemes in the HeCB phase is to vary the core mass undergoing convective mixing.
We refer to $\mbox{{\it convective-core mass}}$ (\Mcc) as the core mass in which $\gradr \geq \grada$ (formal \schw\ core), while the $\mbox{{\it mixed-core mass}}$ (\Mmc) includes also the extra-mixing region (e.g convective core $+$ overshooting and/or semiconvection). The mixed-core mass also indicates the location in the stellar structure of the discontinuity in the chemical composition due to HeCB.
In the ES models (in the  domain we were able to explore) and BS models, \Mcc\ corresponds to \Mmc\, since no extra mixing is introduced. This is also true for the BaSTI-SC model when $Y_{\rm c} > 0.7$, i.e. where semiconvection has not appeared yet, and during the last stages of the HeCB, where the convective core grows rapidly and the size of the semiconvective region is reduced (see Figure \ref{fig:M-L_age}, upper panel).
In HOV, HPC, PARSEC-LOV, and PARSEC-MPC, \Mcc\ and \Mmc\ are distinct from the beginning to the end of the HeCB phase (Figure \ref{fig:M-L_age}, upper and middle panels). Although penetrative convection models are expected to have smaller \Mcc\ and \Mmc\ than overshooting models \citep[see ][ in the case of massive main-sequence stars]{Godart07, Noels_etal10}, max(\Mmc)  in HOV and HPC  is very similar, providing a similar $L_\mathrm{AGBb}$ (see Figure \ref{fig:M-L_age}, upper and lower panels).    

\subsection{Luminosity and duration of the HeCB and eAGB phases}
\label{sec:L&age}
An obvious effect of increasing \Mmc\ is to increase the duration of the HeCB phase ($\tau_\mathrm{HeCB}$). In models with larger \Mmc\, more fuel is available for the \trialfa\ and \coreaction\ nuclear reactions, increasing the time necessary to deplete all the helium in the core, i.e. the lifetime of the HeCB phase. In HOV and HPC models, the HeCB lifetime increases by about $40\%$ compared to the BS model, while in BaSTI-SC and PARSEC-LOV by about $20\%$.
However, if we look at the lifespans of the single phases, the increase in duration of the HeCB phase corresponds to a (non-linearly proportional) decrease of the duration of the AGB phase, since the formation of the He shell takes place closer to the H shell. The BS model has shorter lifetime with respect to the HPC and HOV models, however, it has a longer AGB phase. The BaSTI-SC, PARSEC-LOV, and PARSEC-MPC tracks have a behaviour which is in between the BS and the HPC/HOV models.
In general we expect that models with increasing \Mmc\ produce a longer HeCB phase and a less populated early-AGB, combined with a decreasing ${R_2}$ factor from BS models to the HOV and HPC models. These results are reported in Table \ref{tab:summary}. 
                         
The stellar luminosity is also affected by the mixing scheme adopted during HeCB (Figure \ref{fig:M-L_age}, lower panel), and $L_\mathrm{AGBb}$ increases when extra mixing is added, while it remains almost unchanged when comparing HOV and HPC tracks  (Table \ref{tab:summary}).
The maximum extension of \Mmc\  (which is very similar in HOV and HPC models) corresponds to the inner border of the He-shell at its ignition and it determines the $L_\mathrm{AGBb}$.

Models with larger cores (HOV and HPC)  show \textit{micropulses} \citep{Mazzitelli_DAntona86}, i.e. a series of secular instabilities that may occur during the formation of the He-burning shell. 
Micropulses appear after the maximum in luminosity of the AGB bump if the nuclear reactions in the core stop before the shell is ignited in ``thin'' conditions (see \citealt{Schwarzschild_Harm65}). 
To support this statement we have looked at the central helium abundance and the nuclear energy generation rate at the start of the AGB bump. The values found indicate that in the HPC and HOV models the contribution of HeCB to the luminosity is negligible, since helium is almost completely depleted. On the other hand, in the BS model core-He burning still contributes significantly to the luminosity.
We note that for micropulses to be resolved by a stellar evolution code, a small timestep is needed in the numerical simulations: each pulse lasts for about $5 \cdot 10^{4}$ yr and the duration of the entire phenomenon is about 2 million years.

The study of micropulses goes beyond the purpose of this paper and for detailed explanations we refer the reader to  \citet{Mazzitelli_DAntona86} and the more recent work by \citet{Gautschy&Althaus07}. Also notable of citation are the papers by \citet{Bono_etal97} and \citet{Sweigart_etal00} where the connection between {\it ``gravonuclear instabilities"} and micropulses is investigated.

\begin{table*}
\begin{minipage}{126mm}
\caption{Non-seismic properties of 1.5-\Msun\ models computed adopting different mixing schemes.\label{tab:summary}}
\begin{tabular}{@{}lccccccc}               
\hline                                                           & MESA   & MESA   & MESA   & MESA   & BaSTI  & PARSEC & PARSEC \\
                                                                 & BS      & HOV    & HPC    & ES      & SC     & LOV     & MPC \\
\hline
\hline timespan HeCB phase ($\tau_\mathrm{HeCB}$) [Myr]          & 60.4   & 126.2  & 131.1  & ---    & 100.2  & 100.6  & 107.0  \\
\hline $\mathrm{R_2}$ ratio ($\frac{\tau_\mathrm{AGB}}{\tau_\mathrm{HeCB}}$) & 0.71   & 0.06   & 0.06   & ---    & 0.12   & 0.15   & 0.12   \\
\hline \Mcc\ at $\mbox{$Y_c\sim0.7$ [\Msun]}$                    & 0.100  & 0.113  & 0.084  & 0.162  & 0.165  & 0.140  & 0.107  \\
\hline \Mmc\ at $\mbox{$Y_c\sim0.7$ [\Msun]}$                    & 0.100  & 0.292  & 0.242  & 0.162  & 0.165  & 0.187  & 0.206  \\
\hline maximum extension \Mcc\ [\Msun]                           & 0.100  & 0.130  & 0.114  & ---    & 0.243  & 0.199  & 0.158  \\
\hline maximum extension \Mmc\ [\Msun]                           & 0.100  & 0.305  & 0.302  & ---    & 0.243  & 0.251  & 0.262  \\
\hline
\end{tabular}
\end{minipage}
\end{table*}


\section{Asteroseismic signatures}
\label{sec:sismo}

\subsection{Asymptotic gravity-mode period spacing}
\label{sec:DPg}
We have shown that there are no significant differences between models with similar \Mmc\ when considering the luminosity as a function of time (see e.g. HOV and HPC in Fig. \ref{fig:M-L_age}). 
These models, however, have very distinct seismic properties. \citet{Montalban_etal13} showed that extending the adiabatically stratified central region leads to a larger value of the gravity-mode period spacing.
The reason why the asymptotic period spacing of gravity modes (\dpg) during the HeCB phase is a sensitive probe of the temperature stratification of near-core regions is directly related to the behaviour of the  Brunt-V\"{a}is\"{a}l\"{a} frequency $N$, and its relation with \dpg.

\begin{figure*}
\begin{minipage}{0.42\textwidth}
\centering
\resizebox{\hsize}{!}{\includegraphics{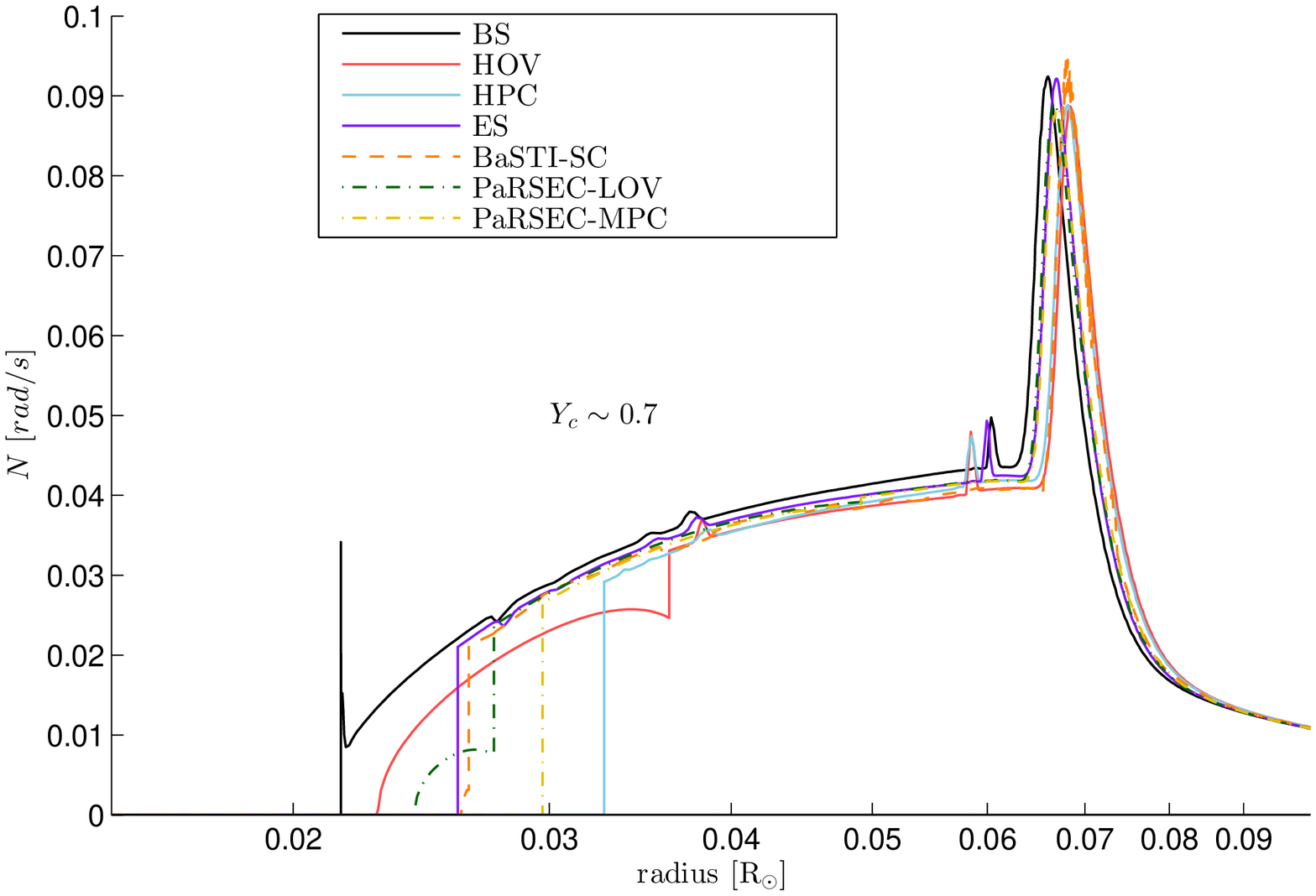}}
\bf{(a)\\}
\resizebox{\hsize}{!}{\includegraphics{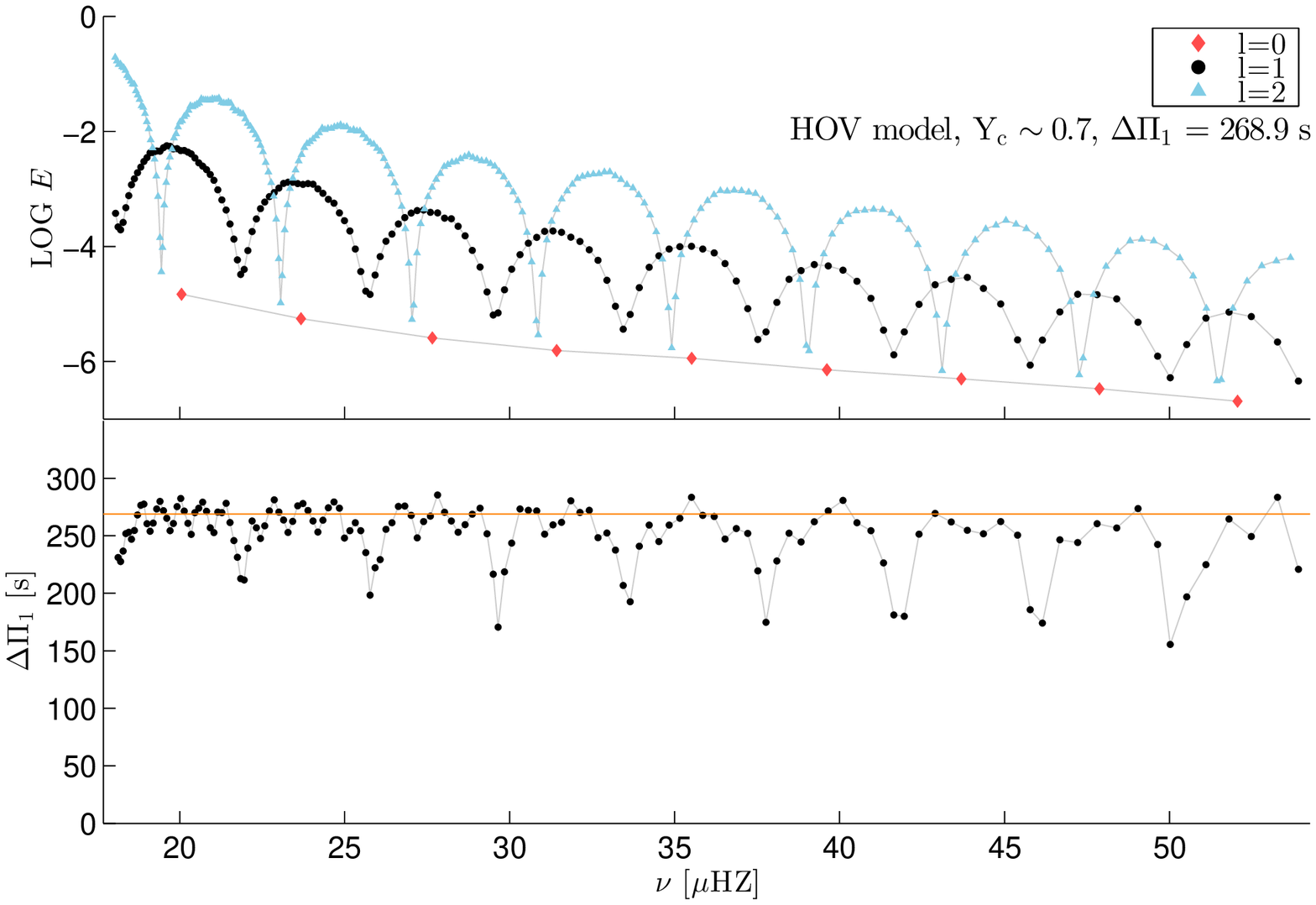}}
\bf{(c)}
\resizebox{\hsize}{!}{\includegraphics{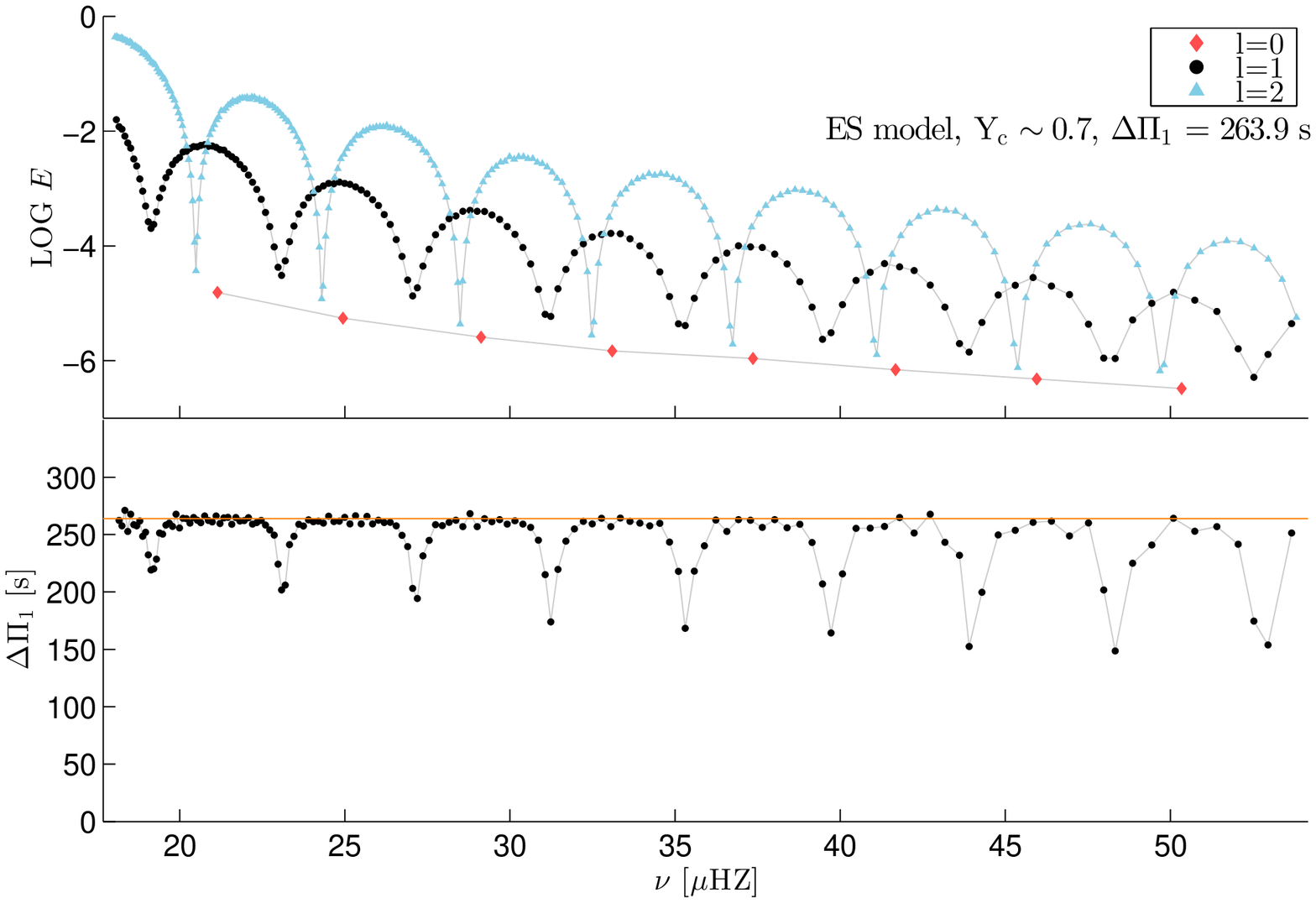}}
\bf{(e)}
\resizebox{\hsize}{!}{\includegraphics{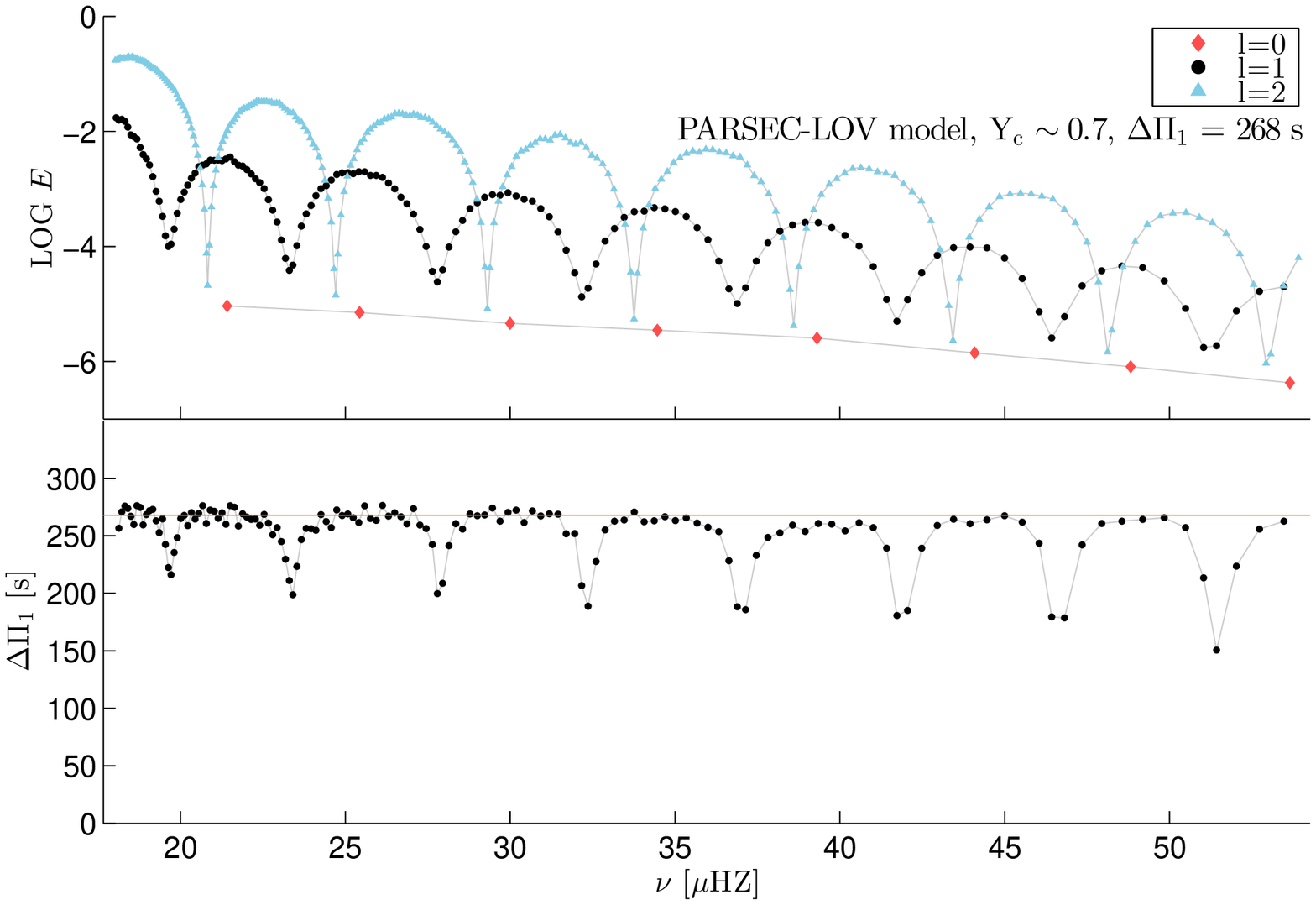}}
\bf{(g)}
\end{minipage}
\begin{minipage}{0.42\textwidth}
\centering  
\resizebox{\hsize}{!}{\includegraphics{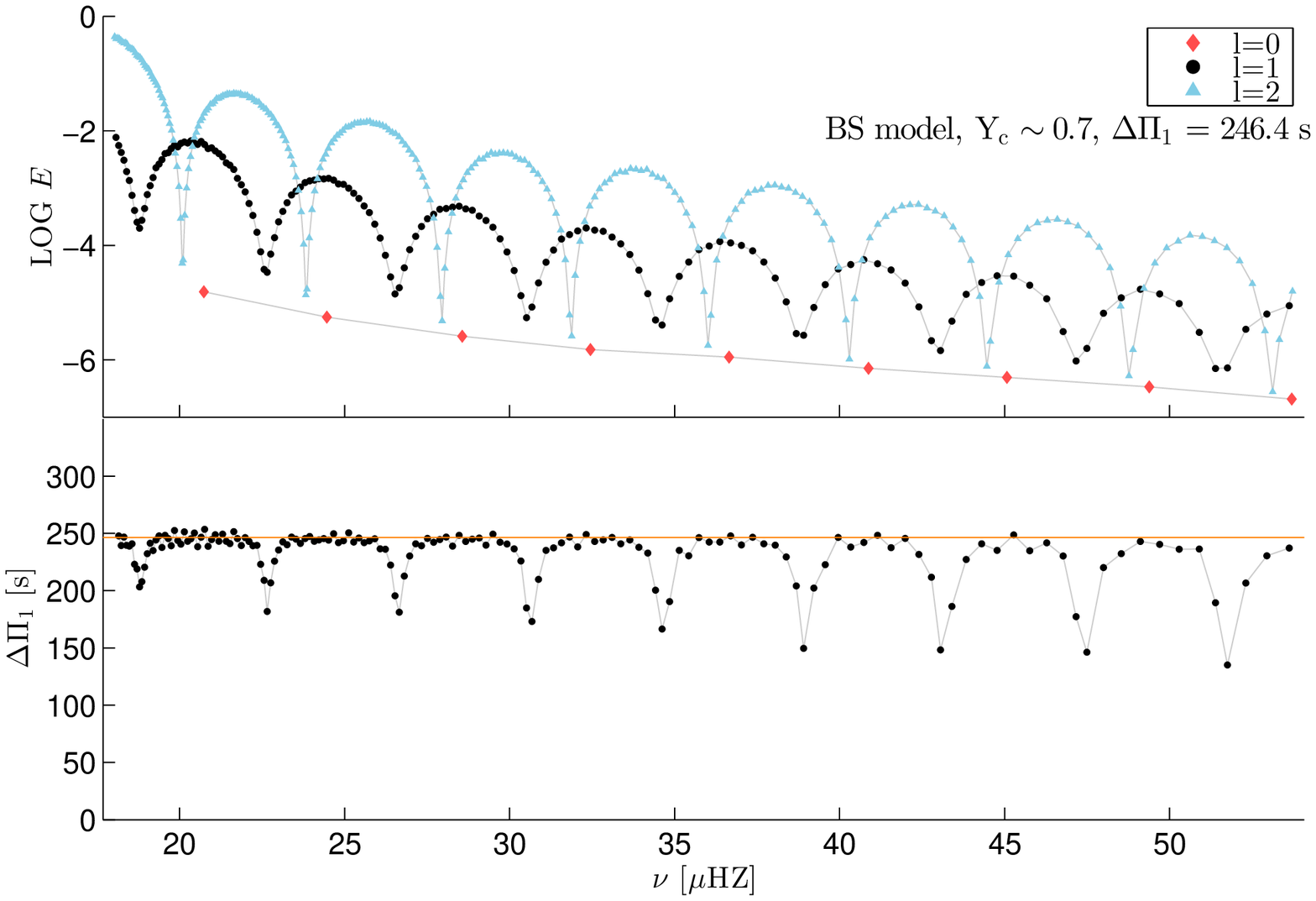}}
\bf{(b)}
\resizebox{\hsize}{!}{\includegraphics{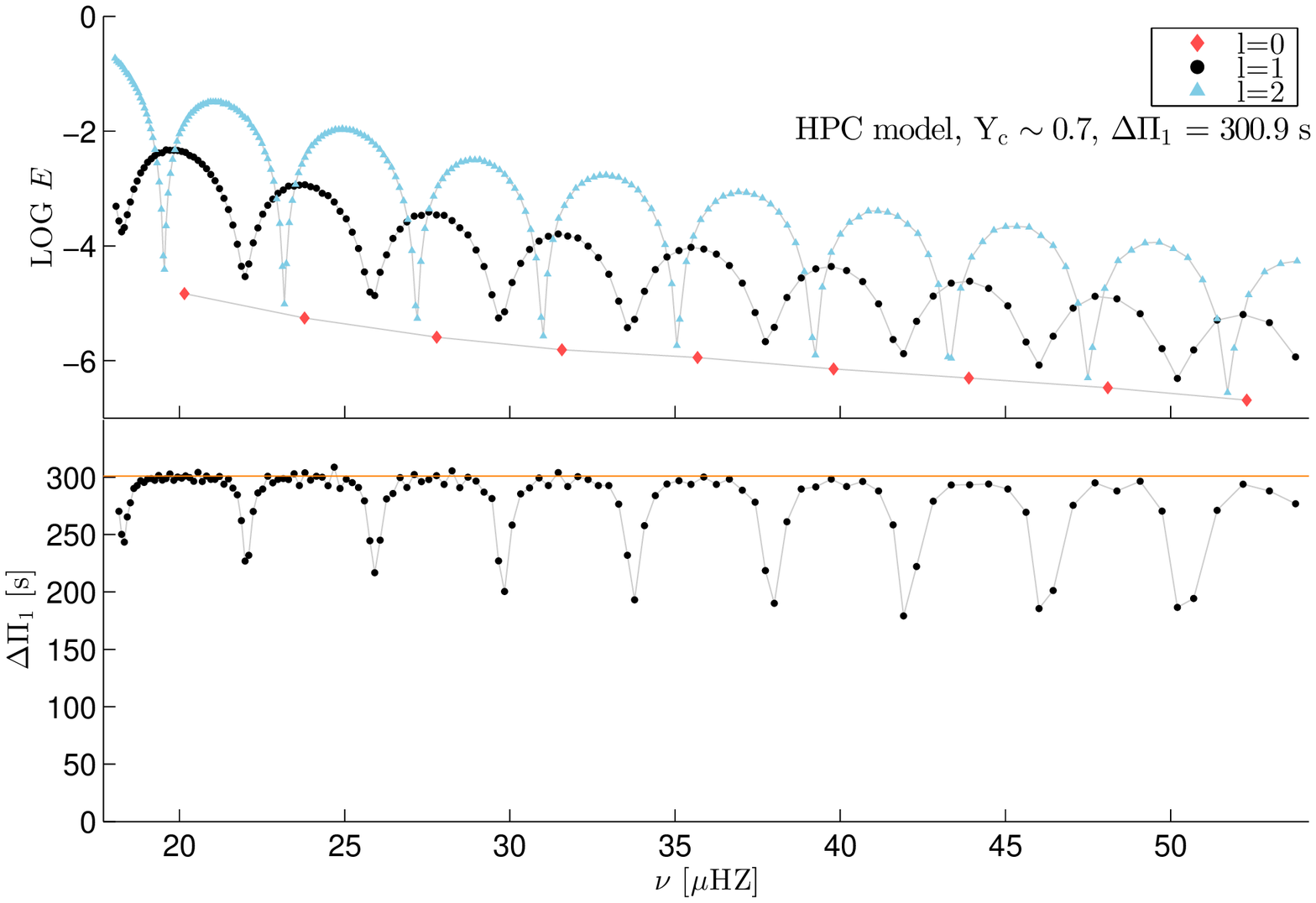}}
\bf{(d)}
\resizebox{\hsize}{!}{\includegraphics{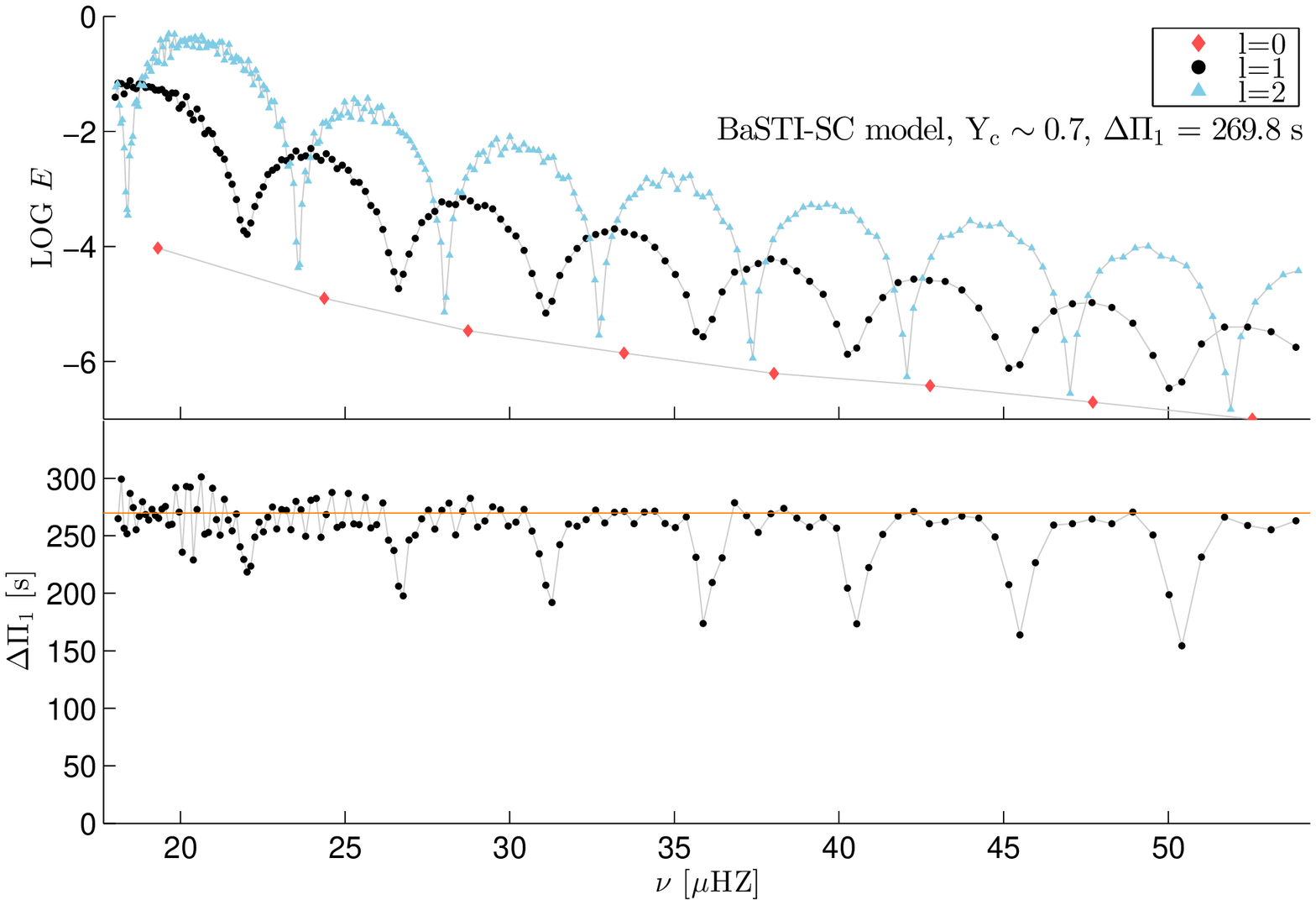}}
\bf{(f)}
\resizebox{\hsize}{!}{\includegraphics{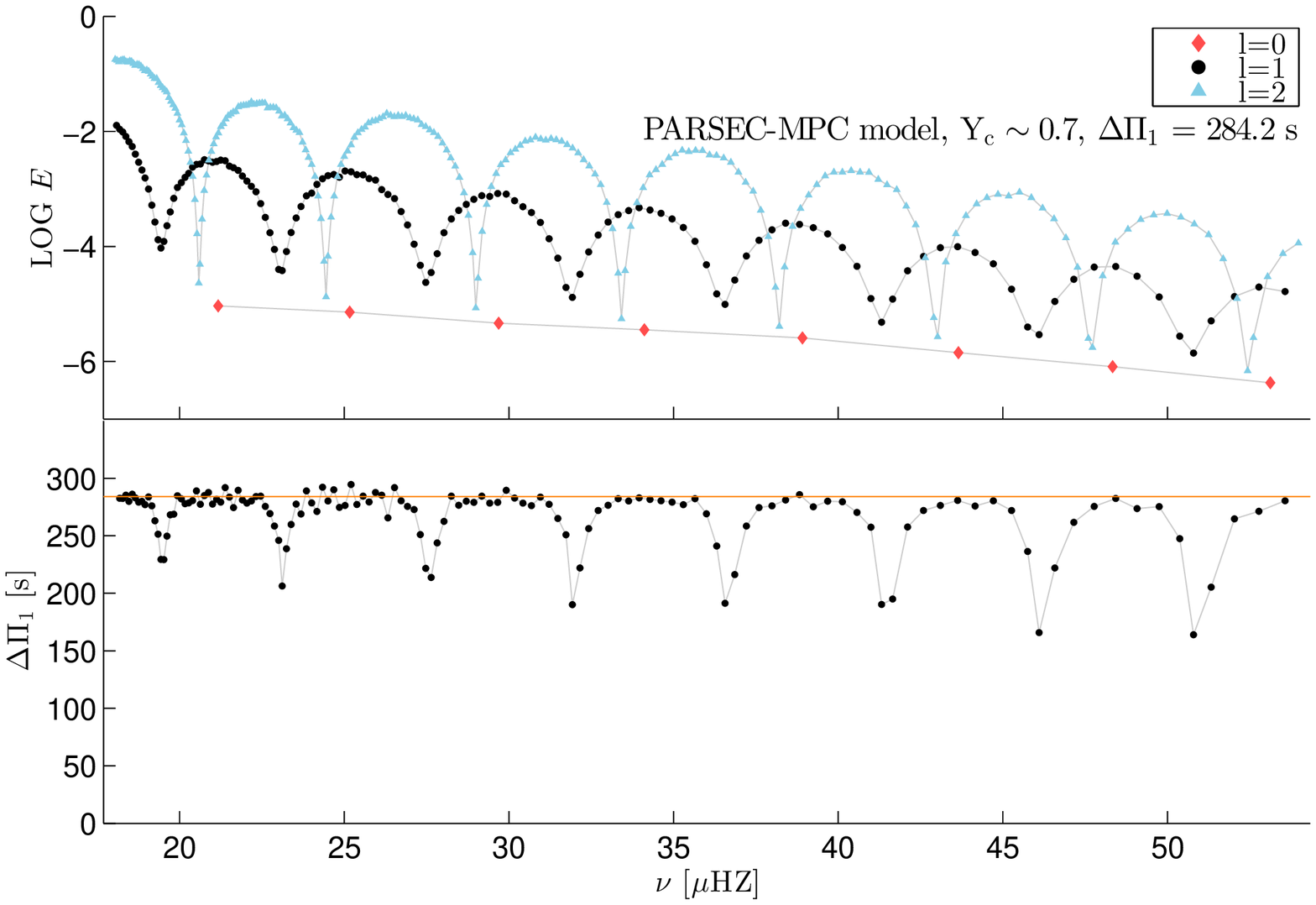}}
\bf{(h)}
\end{minipage}
\caption{ (a): Brunt-V\"{a}is\"{a}l\"{a} frequency in the stellar interior of models with $Y_c\simeq 0.7$ and different convective-mixing scheme.
Figures (b-h): Oscillation modes properties for the different convective schemes. {\it Upper panel:} mode inertia as a function of the frequency for modes with angular degree $\ell=0,1,2$. {\it Lower panel:} period spacing of numerically computed dipolar-mode frequencies (dots) compared with the asymptotic value (solid orange line).}
\label{fig:astero}
\end{figure*}

In the stellar interior, $N$ depends on the local temperature and chemical composition gradients:
\beq
N^2= \frac{g^2\,\rho}{P}\frac{\chi_T}{\chi_\rho}\left(\grada-\gradT-\frac{\chi_\mu}{\chi_T}\gradmu\right){\rm,}
    \label{N_freq_grad}
\eeq
where $\nabla_\mu={\rm d}\ln{\mu}/{\rm d}{\ln P}$,   $\chi_T=(\partial\mathrm{ln}P/\partial\mathrm{ln}T)_{\rho,\mu}$, $\chi_\rho=(\partial\mathrm{ln}P/\partial\mathrm{ln}\rho)_{T,\mu}$, and $\chi_\mu=(\partial\mathrm{ln}P/\partial\mathrm{ln}\mu)_{T,\rho}$.
In a fully mixed region,  \gradmu\ is null, therefore the $N^2$ profile is directly proportional to the difference between \gradT\ and \grada\ (Figure \ref{fig:astero}a).

In the extra-mixed region of models with overshooting (e.g. HOV and PARSEC-LOV) $\gradr < \grada$, therefore $$N^2\propto\grada-\gradT=\grada-\gradr>0\rm ,$$  while in the corresponding region of a penetrative convection model (e.g. HPC and PARSEC-MPC) $\gradT=\grada$, hence $$N^2\propto\grada-\gradT=0 \rm .$$

\begin{figure}
\centering
\resizebox{\hsize}{!}{\includegraphics{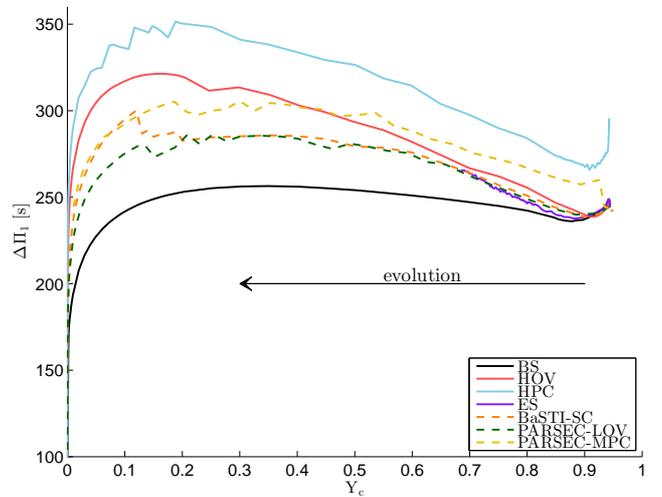}}
\caption{Period spacing against the central helium mass fraction during the HeCB phase for the model computed. Bare-\schw\ models present the smallest period spacing, while the penetrative convection models have the largest values. Moreover penetrative convective models start with larger period spacing.}  
\label{fig:DP-hec}
\end{figure}

The asymptotic period spacing of gravity modes is related to the Brunt-V\"{a}is\"{a}l\"{a} frequency according to the relation \citep{Tassoul80}:
\beq
    \Delta\Pi_{l} = \frac{2\pi^2}{\sqrt{l(l+1)}}
    \left( 
    \int^{r_2}_{r_1}N\frac{\mbox{d$r$}}{r}
    \right)
    ^{-1}{\rm ,}
    \label{eq:Dpg}
\eeq
where $r_{\rm 2}$ and $r_1$ are the boundaries of the g-mode cavity (i.e. where $\omega^2=N^2$). 
Consequently, HeCB penetrative-convection models have higher values of \dpg\ compared to overshooting models (Figure \ref{fig:DP-hec}).

A similar effect can be found if we compare models with increased mixed-core size. Bare-\schw\ models, in fact, have lower \dpg\ compared all  other models, followed by models with induced overshooting (BaSTI-SC and ES), then high overshooting model, and finally high penetrative convection models (see Fig. \ref{fig:astero}a). 

Independently from the convective-mixing scheme adopted (BS, HOV and HPC), the period spacing of models in the HeCB is higher compared to that on the RGB at the same luminosity ($\dpg_\mathrm{RGB}\sim60-50$ s), while after the early-AGB phase \dpg\ decreases to similar or smaller values \citep{Montalban&Noels13}.

It is interesting to notice that if models present multiple gravity-mode cavities, we do not expect them to show a regular period spacing. 
This is the case of HOV and HPC models in the in the post-HeCB phase during the micropulses. In fact, the He shell can experience convection within each pulse, in conjunction with the maximum of the nuclear energy generation.
Stars presenting this scenario (if any at all exist) may be missed by analysis based upon looking for a simple pattern in \dpg\ \citep[e.g.][]{Mosser_etal12}.

\subsection{Period spacing of numerically computed adiabatic frequencies}
From the observed frequency spectra we can estimate \dpg\ based on the detection of modes that have relatively small inertias, i.e. gravity modes that have a significant coupling with the low-inertia acoustic modes (see e.g. \citealt{JCD12} and reference therein).
The frequencies of these mixed gravito-acoustic modes are expected from theory to follow a relatively simple pattern \citep[see e.g.][]{Unno_etal89, Mosser_etal12}, which can be fitted to the observation to estimate \dpg, provided that a sufficient number of modes are detected, and that the analytical approximation for the expected pattern of mixed modes is accurate \citep[see e.g.][]{Beck_etal11, Montalban_etal13, Mosser_etal12, Jiang_JCD14}.

It is thus crucial not only to make predictions of RC \dpg\ using the approximated expression in Eq. \ref{eq:Dpg}, but also to compute the spectrum of individual modes, which may show interesting departures from the expected approximated relation/pattern, and that eventually can be compared with the detailed observed frequency spectrum.

In Figures \ref{fig:astero}b$-$f we present the properties of adiabatic pulsation modes computed with GYRE \citep{GYRE}, and compare  frequency spectra of models with the same central helium abundance ($Y_{\rm c}\simeq 0.7$), but computed with different convective-mixing schemes. In the upper panel of each figure it is possible to see how the inertia ($E$) of each mode varies in a frequency interval around the frequency of maximum oscillations power (\numax).
Figures \ref{fig:astero}b$-$f also show that the characteristic asymptotic behaviour of the modes (the constant frequency separation for the low-inertia,  pressure-dominated modes and the constant period spacing for the high-inertia, gravity-dominated modes) is a good representation of the detailed, numerically computed frequency spectrum. Moreover,  the asymptotic value of the period spacing clearly reflects the differences in the Brunt-V\"{a}is\"{a}l\"{a} frequency near the core (Fig. \ref{fig:astero}a and Eq. \ref{eq:Dpg}), with the BS model having the lowest \dpg, and the HPC model the highest.  

\subsection{Signatures of sharp-structure variations in the period spacing}
\label{sec:glitch}

\begin{figure}
\resizebox{\hsize}{!}{\includegraphics{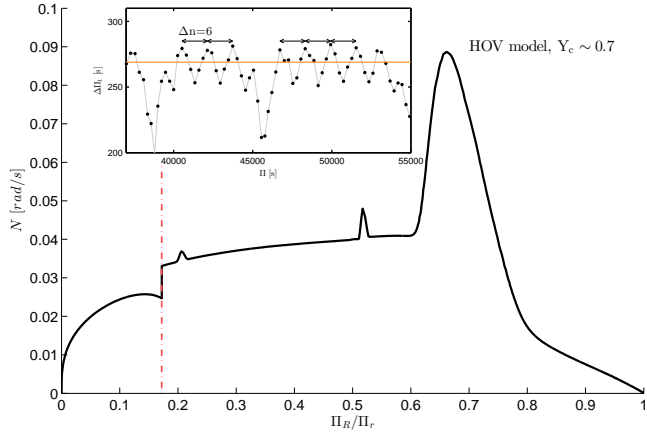}}
\caption{
Brunt-V\"ais\"al\"a frequency as a function of the normalised buoyancy radius ${\Pi_{\rm r}}/{\Pi_R}$ (see text) in a HOV model with $Y_{\rm c} \simeq 0.7$. In the rectangle, $\ell=1$  period spacing as a function of period for the same model. 
The period spacing of high-order g modes (e.g. in the period range $4.7-5.3$  $10^4$ s)  is well described by the superposition of the asymptotic \dpg\ (orange line) and a component with periodicity $\Delta n\simeq 6$. This periodicity indicates  (see Eq. \ref{eq:variak}) a sharp-structure variation located at ${\Pi_{R}}/{\Pi_{\rm r}}\simeq1/\Delta n\simeq 0.17$, which corresponds well with the position of glitch in $N$.}
\label{fig:BVglitch}
\end{figure}

As evinced from Fig. \ref{fig:astero}a and, more clearly, from 
Fig. \ref{fig:BVglitch}, $N$ may have sharp changes due e.g. to chemical composition gradients  and/or in the temperature gradient in radiative regions.  Whether such glitches have a significant impact of the period spacing depends on their location, their sharpness, and to the typical local wavelength of the gravity modes of interest.

As described in the literature  \citep[see e.g.][]{Brassard_etal92, Miglio_etal08,Berthomieu&Provost88},  the signature of a sharp feature in the Brunt-V\"ais\"al\"a frequency is a periodic component in the periods of oscillations, and therefore  in the period spacing, with a periodicity in terms of the radial order $n$ given by:
\begin{equation}
\Delta n\simeq\frac{\Pi_R}{\Pi_{\rm glitch}}\;{,}
\label{eq:variak}
\end{equation}
where  the  total buoyancy radius is defined as:
\begin{equation}
\Pi_R^{-1}=\int_{r_0}^R{\frac{N}{r'}dr'}\;{\rm ,}
\end{equation}
and local  buoyancy radius is
\begin{equation}\Pi_r^{-1}=\int_{r_0}^r{\frac{N}{r'}dr'}\;{\rm ,}
\end{equation}
with $r_0$  and $R$ being the inner and outer boundary of the g-mode propagation region.
The periodicity of the components in terms of radial order is therefore an indicator of the location of the glitch, expressed in terms of its normalised buoyancy radius (${\Pi_R}/{\Pi_{\rm glitch}}$). 

In the models we are focusing on, we notice two main glitches. 
A prominent, yet smooth, glitch due to the H-burning shell and associated with $\nabla_\mu$ (see e.g. the glitch located at ${\Pi_R}/{\Pi_{\rm r}} \simeq 0.7$ in 
Fig. \ref{fig:BVglitch}). In models at the beginning of the HeCB phase, in which the H-burning shell is still very thin, this glitch is sharper and may give rise to significant departures from a smooth g-mode period spacing (see Bildsten et al. in preparation). In models during most of the HeCB phase, however, this glitch does not appear to give rise to significant deviations from the asymptotic \dpg\ expected for high-order g modes. 

More interestingly, in the HOV model (see Fig.  \ref{fig:astero}c  and \ref{fig:BVglitch}) we notice a sharp variation in $N$  which can be well described by a step function\footnote{At the boundary of the fully mixed radiative (overshoot) region, the discontinuity in opacity, due to the difference between carbon rich mixed layers and He rich surrounding layers, leads to a discontinuity in \gradT\ and thus in $N$.}. 
Given the location of this glitch $\Pi_R/\Pi_r \simeq 0.17$ (see Fig. \ref{fig:BVglitch}) we  expect a periodic deviation from the asymptotic \dpg\ with a periodicity of  $\Delta n \simeq 6$, where $n$ is the radial order of gravity modes (see e.g. \citealt{Miglio_etal08}). This corresponds well (at least in the regions of pure g modes, e.g. in the range $4.7-5.3$  $10^4$ s) to the periodicity of the component (see inset of  Fig. \ref{fig:BVglitch}).

While the full description of these glitches is beyond the scope of the present paper, we note that departures from the simple description of \dpg\ expected from the interaction between high-order g modes and an acoustic mode \citep[e.g., see]{Unno_etal89, Mosser_etal12, Jiang_JCD14} provide  additional, potentially very sensitive, probes of sharp-structure variations in near-core regions during the HeCB phase.

\begin{figure}
\centering
\resizebox{\hsize}{!}{\includegraphics{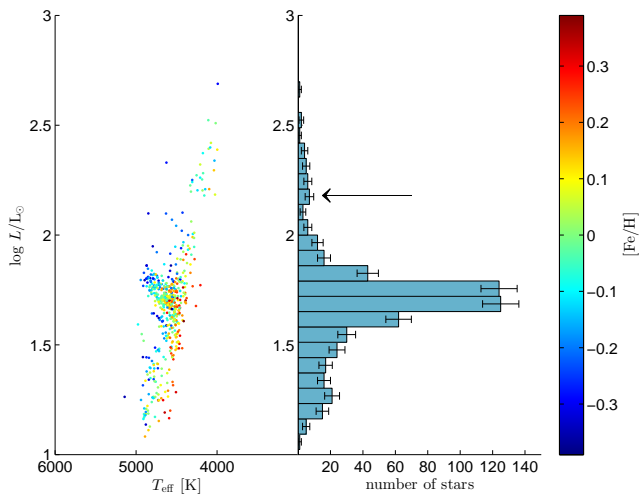}}
\caption{Hertzsprung-Russell diagram and luminosity distribution of the APOKASC catalogue \citep{APOKASC}. The sample is selected in a range of mass \mbox{$M=$1.3--1.7\Msun} and metallicity \mbox{$[\mathrm{M/H}]=$-0.4--0.4}. The AGB bump can be identified in the right the panel at around $\logL\sim2.2$ (black arrow) while the major peak correspond to the RC.}
\label{fig:apokasc}
\end{figure}

\section{First comparison with observations: AGB bump and period spacing}
\label{sec:data}

\subsection{The AGB bump in  \kepler\ red giants }
A catalogue of seismic  (\dnu\ and \numax) and spectroscopic  ([M/H] and \Teff) constraints for $\sim 1600$ \kepler\ giants was recently published by the APOKASC collaboration \citep{APOKASC}. 

We estimate stellar masses, and luminosities of these stars by using the so-called direct method, i.e. by combining $T_{\rm eff}$ with the seismic radii (estimated using \dnu\ and \numax\ through scaling relations). This method is known to lead to less precise estimates of $M$ and $R$ than so-called ``grid-based'' approaches, yet it is less dependent on stellar evolutionary tracks \citep[e.g. see][ and references therein]{ChaplinMiglio_13} and completely independent of bolometric corrections.
We select stars from the APOKASC catalogue in a range of mass \mbox{$M=$1.3--1.7 \Msun} and metallicity \mbox{$[\mathrm{M/H}]=$-0.4--0.4}. 
The luminosity function of such stars displays a peak that is spread over about 4 bins and a has maximum at $\logL\sim2.2$: we interpret this peak as a strong candidate for the AGB bump (Figure \ref{fig:apokasc}).
This statement is supported by the fact that $97.5\%$ of the stars have errors smaller than the bin size. 
We also calculated 1000 realisations of the observed sample, assuming gaussian errors on \dnu, \numax, \Teff\ (taken from the APOKASC catalogue) and found that the properties of the peak in the luminosity function are not significantly affected.

As a word of caution we would like to stress that the APOKASC catalogue may, however, be affected by target selection biases \citep{APOKASC}.
Although the maximum in the observed luminosity distribution  does not appear to be significantly affected by widening the metallicity range, we notice that, if we extend the range of masses down to 1 \Msun, the position of the peak is lowered by 1 bin (0.07 dex).


\begin{figure}
\centering
\resizebox{\hsize}{!}{\includegraphics{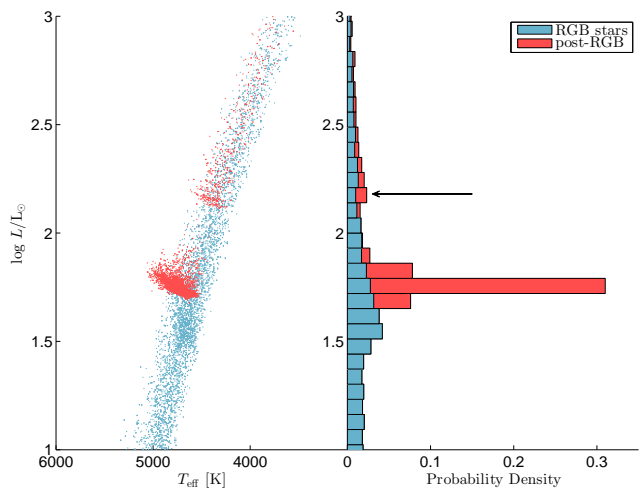}}
	\caption{Hertzsprung-Russell diagram (left panel) and luminosity density distribution (stacked histogram, right panel) of stars in the TRILEGAL simulation of the {\it Kepler} field for the same masses and metallicities as in  Figure \ref{fig:apokasc}. The AGB bump is visible in the right panel around $\logL\sim2.2$ (black arrow). }
\label{fig:hist_trilegal}
\end{figure}

To check whether the AGBb is a feature we expect to be able to detect in a composite stellar population we use the TRILEGAL code \citep{Girardi_etal12} to simulate the galactic population expected in {\it Kepler} field.  The stellar models used are based on Padova tracks \citep{Bressan_etal12}, with an overshooting parameter of $\Lambda_c=0.5$ (same mixing scheme adopted for PARSEC-LOV model). We apply to the synthetic sample the same selection in mass and metallicity  as in the observed sample.
The synthetic population also displays a well defined peak with a luminosity compatible with that of the candidate AGBb in the APOKASC catalogue (Figure \ref{fig:hist_trilegal}). 

From the simulations we estimate that about half of the red-giant stars in that peak belong to the RGB. This means that in the case of the full  APOKASC catalogue (considering the entire range of mass and metallicity) we expect about $20-40$ AGBb stars. 

\begin{figure*}
\centering
\begin{minipage}{\textwidth}
\centering
\resizebox{0.8\hsize}{!}{\includegraphics{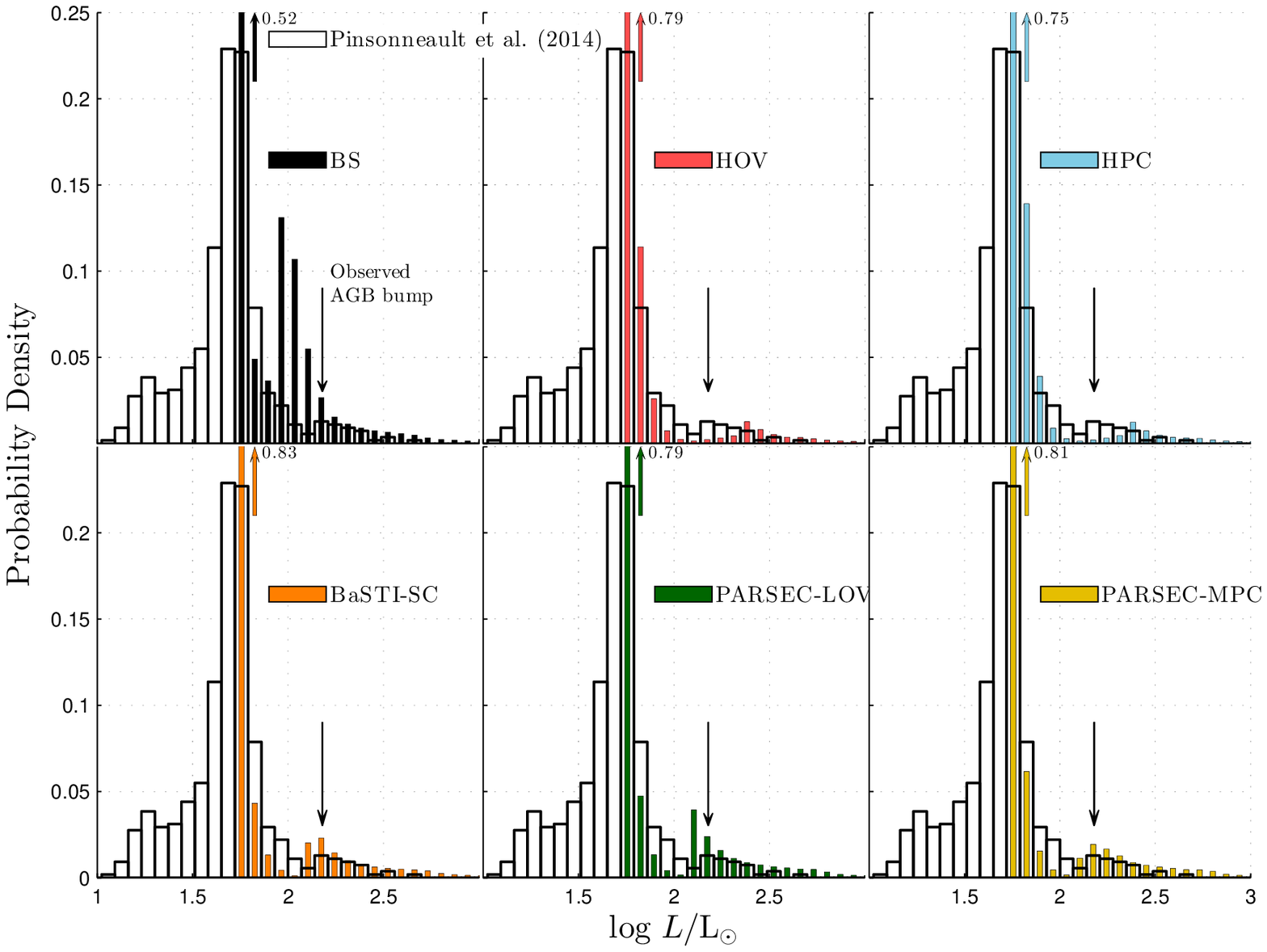}}
\caption{Comparison between the observed luminosity density distribution of stars in APOKASC public catalogue selected with the same criteria as in Figure \ref{fig:apokasc}, and the predictions from the models described in the paper. In the models, RGB stars are omitted.}
\label{fig:hist_popL}
\resizebox{0.8\hsize}{!}{\includegraphics{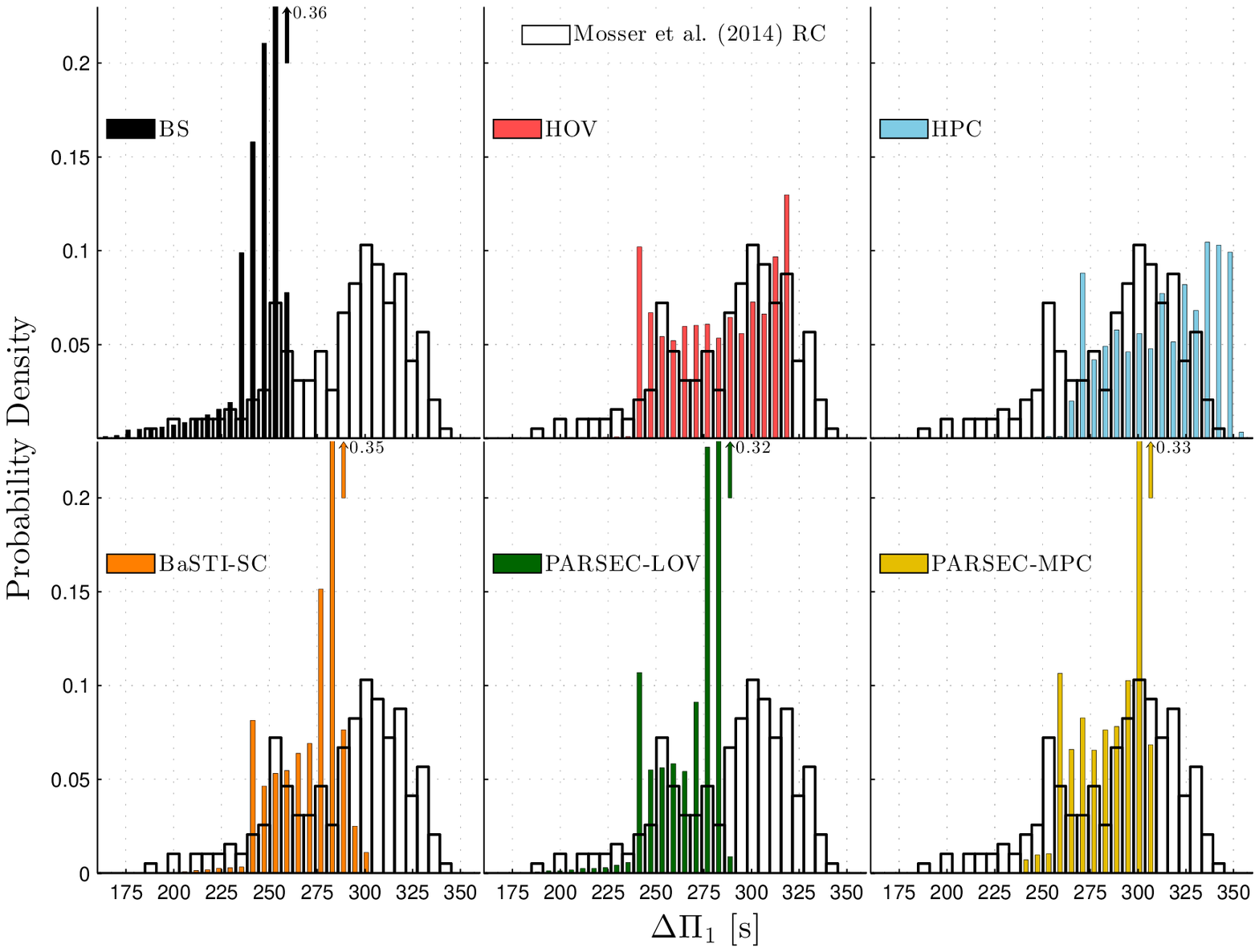}}
\caption{Comparison between the observed period-spacing distribution of stars classified by \citet{Mosser_etal14} as RC and AGB and the predictions from the models described in the paper. We have selected stars with mass $1.3 \leq M/{\rm M_\odot} \leq 1.7$.}
\label{fig:hist_popPg}
\end{minipage}
\end{figure*}

In order to make a first comparison between our models and the  observations, we generate histograms of the luminosity based on each of our tracks.
Results of this comparison are shown in Figure \ref{fig:hist_popL}. For the sake of clarity in the figure we have omitted RGB models. All the models are able to reproduce reasonably well the position of the RC, taking into account also the fact that our models are representative of a single-mass, single-metallicity population only, and that we have not added the effect of observational uncertainties when building the synthetic luminosity function.

The luminosity of the AGB-bump predicted by BaSTI-SC, and PARSEC-MPC  are remarkably similar, and in good agreement with the candidate AGB-bump luminosity as detected in the observations. While PARSEC-LOV is another acceptable model, interestingly, the AGBb predicted by the BS, HOV, and HPC models is in clear disagreement with the observations, being either too faint (BS) or too bright (HOV, HPC).  

\subsection{Period spacing of \kepler\, RC stars}
\citet{Mosser_etal14} recently published a catalogue that contains a large sample of red giants in the \kepler\ field for which g-mode asymptotic period spacings were inferred by modelling the interaction between pressure and gravity modes \citep[see ][]{Unno_etal89, Mosser_etal12}. Thanks to this catalogue we are able to compare the observed period spacing with theoretical predictions. We select RC stars  (i.e. stars with $\dpg>175$ s) in a range of mass between 1.3 \Msun\ and 1.7 \Msun, and compare the distribution of observed \dpg\ with predictions from our models (see Fig. \ref{fig:hist_popPg}). 

None of the models considered seems to describe satisfactorily the entire observed distribution. The BS model can be  ruled out, since $\max{\dpg}$ is $80-90$ s smaller than the maximum in the observed data. 
BASTI-SC, which was one of our best candidate models based on the comparison with the AGBb luminosity, shows a main peak in the  \dpg\ distribution about  $20-30$s below the observed one and cannot describe period spacings higher than $300$s .
PARSEC-LOV has a very similar behaviour to BaSTI-SC and has a maximum $\dpg\simeq285$ s.
The HPC model is able to reproduce a large domain of the observed distribution, although its \dpg\ appears shifted to higher values with respect to the observations.
Finally, although PARSEC-MPC and HOV cannot reproduce the highest observed  \dpg, their main peaks agrees with the observed one within 1-2 bins resolution.  

In the observed distribution of \dpg (see Fig. \ref{fig:hist_popPg})  we notice, in addition to the main peak, a secondary peak at $\dpg \sim 253$ s, which still belongs to the RC population.
Interestingly all our models, but BS, show the presence of this second structure. The reason behind it can be deduced simply following the theoretical tracks in the $Y_{\rm c}-\dpg$ diagram (see Fig. \ref{fig:DP-hec}). The tracks cross 3 times a narrow interval of period spacing: the first two times at the very beginning of the HeCB phase, i.e where \dpg\ decreases until a local minimum and then starts to rise, while the third passage happens during the rapidly decrease that follows the absolute maximum (at the end of the HeCB). Since time and $Y_{\rm c}$ are, in first approximation, linearly related during the HeCB, the third passage is very quick compared to the first and the second, therefore the peak is almost entirely populated by stars at the beginning of the phase.  In the BS model the main and the second peak are not resolved since \dpg\ stays nearly constant for most of the HeCB phase.

At this stage we are however limited to a qualitative comparison between observed and theoretical distributions of \dpg. Such limitation arises from the fact that observational biases at the target selection stage, and in the determination of the \dpg\ from the power spectra, have not been fully explored yet. 
Moreover, although we have chosen models with a mass and metallicity representative of the stars observed by \kepler\ ($[M/H]=-0.07\pm0.24$),
and selected stars according to the mass, our synthetic population is rather simplistic. As a first test of the effect of changing the mass and metallicity, we consider HOV models  with different mass ($M= 1\ \Msun$) or metallicity ($Z=0.007$). We find that while the  \dpg\  distribution depends little on the mass,  the effect of reducing significantly the metallicity is to extend to the range of \dpg\ by $\sim 20 $ s. This effect needs to be taken into account when making quantitative comparisons between the observed and theoretically predicted distributions.

While these limitations can be partly mitigated by a more realistic synthetic population, and by a thorough examination of selection biases, it is likely that the most robust inference will be possible when applying our test to simple stellar populations, i.e. to red giants in the old-open clusters NGC6791 and NGC6819.

\section{Summary and future prospects}
\label{sec:conclusion}
The ability to predict accurately the properties of He-core-burning stars depends on our understanding of convection, which remains one of the key-open questions in stellar modelling (e.g. see \citealt{Castellani_etal71a,Chiosi07,Salaris07,Bressan_etal15}).  Crucially, stringent tests of models have been limited so far by the lack of observational constraints specific to the internal structure of evolved stars. 

In this work we propose a way forward. We argue that the combination of two observational constraints, i.e. the luminosity of the AGB bump and the RC period spacing of gravity modes,  provides us with a decisive test to discriminate between competing models of HeCB low-mass stars.

We have computed  a series of stellar models with various prescriptions for the transport of chemicals and for the thermal stratification of near-core regions, and using different  evolution codes (MESA, BaSTI, and PARSEC).  First, we used these models to make predictions about the duration of the RC and eAGB phases, the luminosity of the AGB bump, and C/O ratios in WDs. A summary of the models characteristics can be found in Table \ref{tab:summary}.  We then focussed on the prediction of seismic observables (see Sec. \ref{sec:sismo}). We found that the asymptotic period spacing of gravity modes depends strongly on the prescription adopted (with differences up to about 40\%, when comparing the BS and HOV model). 

We complemented this analysis by a numerical computation of adiabatic oscillation frequencies. This allowed us to confirm that the asymptotic approximation (Eq. \ref{eq:Dpg}) is a good representation of the period spacing of gravity-dominated modes. Moreover (see Sec. \ref{sec:glitch}), the detailed behaviour of the period spacing of g modes shows the seismic signature of sharp variations in the Brunt-V\"ais\"al\"a frequency, which could potentially give additional information about near-core features (localised chemical composition gradients and near-discontinuities in the temperature gradient).

We then presented (Sec. \ref{sec:data}) a first comparison between our predictions and the observational constraints obtained from the analysis of \kepler\ light curves \citep{APOKASC, Mosser_etal14}.
We found evidence for the AGB-bump among \kepler\ targets, which allowed us to make a first combined analysis of classical (AGBb luminosity) and seismic (RC \dpg) constrains. 
Our main conclusion is that, while standard models (BaSTI-SC, PARSEC-LOV) are able to reproduce the luminosity of the AGBb, they cannot describe satisfactorily the distribution of the observed period spacing of RC stars while models with high overshooting (HOV), although giving a much better description of the observed RC \dpg\ distribution, fail to reproduce the AGBb luminosity. We then suggest a candidate model to describe simultaneously the two observed distributions: a model with a moderate overshooting region in which we apply an adiabatic thermal stratification. This prescription (which we have tested using PARSEC, see PARSEC-MPC model) gives indeed a better description of the observations. 

At this stage of the analysis we are however prevented from drawing any further quantitative conclusions.
To achieve the latter, we will follow two complementary approaches.
On the one hand, we will couple our models with TRILEGAL, to generate synthetic stellar populations which can be quantitatively compared with the observed composite disk population. We will also investigate in detail possible observational biases, both in the target selection and in the detection of the period spacing from oscillations spectra.
On the other hand,  to limit/quantify such biases, we will test our models considering stars in the clusters NGC6791 and NGC6819, in which oscillations were detected in $\sim 30$  HeCB giants \citep[e.g., see][]{Stello_etal11, Miglio_etal12, Corsaro_etal12}.

\section*{Acknowledgments}
AM acknowledges the support of the UK Science and Technology Facilities Council (STFC). Funding for the Stellar Astrophysics Centre is provided by The Danish National Research Foundation (Grant agreement no.: DNRF106). The research leading to these results has received funding from the European Community's Seventh Framework Programme ([FP7/2007-2013]) under grant agreement no. 312844 (SPACEINN).
AM and LG acknowledge support from PRIN INAF 2014 (PI: L. Girardi) -- CRA 1.05.01.94.05. 
SC is funded by PRIN-INAF 2014 (PI: S. Cassisi) and by the Economy and Competitiveness Ministry of the Kingdom of Spain (Grant AYA2013-42781P).
JM and PM acknowledge the support from the ERC Consolidator Grant funding scheme ({\em project STARKEY}, G.A. n. 615604).
AB acknowledges support from PRIN INAF 2014 "Star formation and evolution in galactic nuclei"


\bibliographystyle{mnras}
\bibliography{bib}


\appendix

\section{Carbon and Oxygen Content}
\label{sec:CO}
As mentioned in Sec. \ref{sec:review}, an additional consequence of adopting different mixing schemes in the HeCB phase is the resulting chemical profile of the C-O core, which affects the  chemical profile of a WD at the beginning of the cooling sequence.

\begin{figure}
\centering
\resizebox{\hsize}{!}{\includegraphics{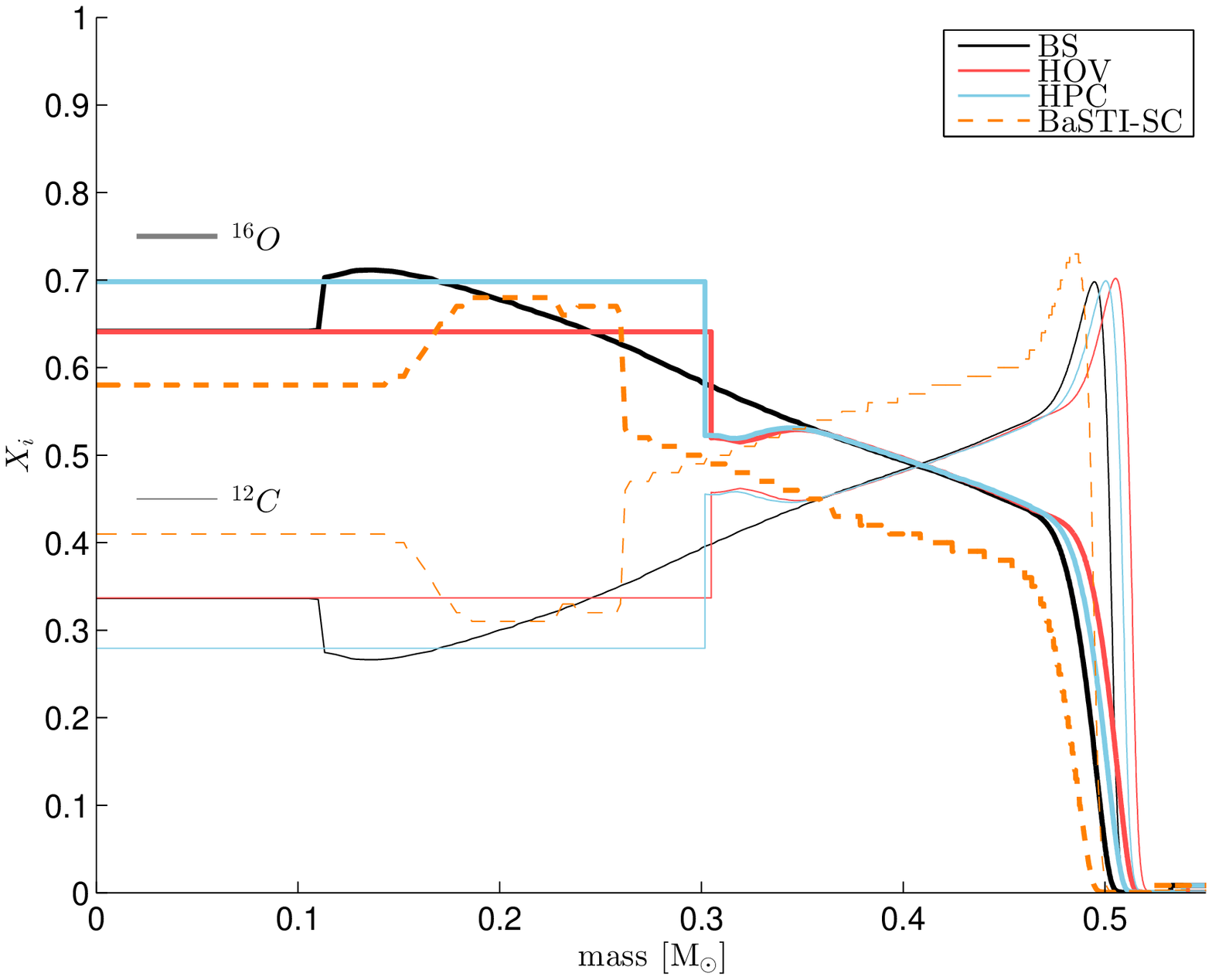}}
\caption{Carbon and Oxygen profile of 1.5-$\Msun$ models at the first AGB-TP.}

\label{fig:CO}
\end{figure}

To provide the reader with a complete set of predictions, we show in Figure \ref{fig:CO} the chemical-composition profile of AGB models at the first thermal pulse.
The results are consistent with what \citet{Straniero_etal03} found in models of a 3 \Msun\ star. They showed that the maximum of $X_\mathrm{O}$ for a bare-\schw\ and induced overshooting (plus semiconvection) models lies outside the former convective core while in models with high overshooting it corresponds to the value found in the former mixed core. 
Moreover, in our HPC model the $X_\mathrm{O}$ is higher than the HOV model, mostly due to the larger duration of the he-burning core phase (figure \ref{fig:CO}, left panel) in which \coreaction\ reaction has more time to convert carbon into oxygen.

\bsp	
\label{lastpage}
\end{document}

%% file: journalDefs.tex
\def\aj{AJ}%
\def\araa{ARA\&A}%
\def\apj{ApJ}%
\def\apjl{ApJ}%
\def\apjs{ApJS}%
\def\ao{Appl.~Opt.}%
\def\apss{Ap\&SS}%
\def\aap{A\&A}%
\def\mnras{MNRAS}%
\def\solphys{Sol.~Phys.}%
\def\memsai{Mem.~Soc.~Astron.~Italiana}%